\documentclass{article}

\usepackage{arxiv}

\usepackage[utf8]{inputenc} % allow utf-8 input
\usepackage[T1]{fontenc}    % use 8-bit T1 fonts
\usepackage{hyperref}       % hyperlinks
\usepackage{url}            % simple URL typesetting
\usepackage{booktabs}       % professional-quality tables
\usepackage{amsfonts}       % blackboard math symbols
\usepackage{nicefrac}       % compact symbols for 1/2, etc.
\usepackage{microtype}      % microtypography
\usepackage{lipsum}
\usepackage{graphicx}
\graphicspath{ {./images/} }
\usepackage{cite}
\usepackage{amsmath,amssymb,amsfonts}
\usepackage{algorithmic}
\usepackage{graphicx}
\usepackage{textcomp}
\usepackage{xcolor}
\usepackage{listings}
\def\BibTeX{{\rm B\kern-.05em{\sc i\kern-.025em b}\kern-.08em
    T\kern-.1667em\lower.7ex\hbox{E}\kern-.125emX}}

\title{Smells-sus: Sustainability Smells in IaC}

\author{
 Seif Kosbar \\
  Software and Emerging Technologies Lab (SAET)\\
  Polytechnique Montréal\\
  Montréal, Canada \\
  \texttt{\href{mailto:seifeldin.kosbar@polymtl.ca}{seifeldin.kosbar@polymtl.ca}}\\
   \And
 Mohammad Hamdaqa \\
  Software and Emerging Technologies Lab (SAET)\\
  Polytechnique Montréal\\
  Montréal, Canada \\
  \texttt{\href{mailto:mhamdaqa@polymtl.ca}{mhamdaqa@polymtl.ca}}\\
}
\date{}

\begin{document}
\maketitle
\begin{abstract}
Practitioners use Infrastructure as Code (IaC) scripts to efficiently configure IT infrastructures through machine-readable definition files. However, during the development of these scripts, some code patterns or deployment  choices may lead to sustainability issues like inefficient resource utilization or redundant provisioning for example. We call this type of patterns sustainability smells. These inefficiencies pose significant environmental and financial challenges, given the growing scale of cloud computing. This research focuses on Terraform, a widely adopted IaC tool. Our study involves defining seven sustainability smells and validating them through a survey with 19 IaC practitioners. We utilized a dataset of 28,327 Terraform scripts from 395 open-source repositories. We performed a detailed qualitative analysis of a randomly sampled 1,860 Terraform scripts from the original dataset to identify code patterns that correspond to the sustainability smells and used the other 26,467 Terraform scripts to study the prevalence of the defined sustainability smells. Our results indicate varying prevalence rates of these smells across the dataset. The most prevalent smell is Monolithic Infrastructure, which appears in 9.67\% of the scripts. Additionally, our findings highlight the complexity of conducting root cause analysis for sustainability issues, as these smells often arise from a confluence of script structures, configuration choices, and deployment contexts.
\end{abstract}

% keywords can be removed
% \keywords{Cloud Computing \and Sustainability \and Infrastructure as Code \and Energy Efficiency.}

\section{Introduction}
Cloud computing is one of the most influential technologies of the 21st century, enabling unprecedented levels of scalability, reliability, and accessibility for various applications and services over the internet. However, cloud computing also comes with a hefty environmental cost, as it consumes enormous amounts of energy. According to the International Energy Agency (IEA), in 2022, data center power consumption reached values close to 240-340 TWh. It is about 1-1.3\% of global energy demand \cite{Iea_2024}. According to another study, by 2025, cloud services are expected to consume 20\% of global electricity and emit up to 5.5\% of the world’s carbon emissions \cite{buyya2023energyefficiency}. Infrastructure as Code (IaC) is the practice of automatically defining and managing network and system configurations, and infrastructure through source code \cite{humble2010continuous}. A study conducted in 2019 investigated the state of practice in IaC adoption based on data from 44 semi-structured interviews in various companies. The study shows that 69\% of the practitioners work with and develop solutions using more than three IaC tools, with 34.1\% of them using Terraform. They prefer using machine-readable definition files like IaC instead of manual configuration tools \cite{8919181}. In this study, we focus on Terraform, one of the most widely used IaC tools, which supports declarative code and multiple major cloud providers like Amazon Web Services (AWS), Microsoft Azure, and Google Cloud Platform (GCP) \cite{terraform, Amazon_Web_Services_2023, GCP, azure}. Terraform scripts enable you to describe various aspects of the cloud infrastructure in human-readable configuration files \cite{terraform}. Cloud configurations can be the number and type of virtual machines (VMs), the network topology, the security policies, and the software dependencies \cite{gcpDeploymentManager}. Terraform scripts can also enable the dynamic scaling and adaptation of the cloud infrastructure according to the workload and performance requirements \cite{hashiCorpTerraform}. IaC scripts, much like traditional software code, are prone to defects, which can lead to serious issues when scaled \cite{Jiang2015}. For instance, a defective IaC script caused significant data loss by deleting home directories for around 270 Wikimedia Commons users in cloud environments in 2017 \cite{commons2017incident}. Another case involved a flawed IaC script triggering an Amazon Web Services (AWS) outage, resulting in financial damages estimated at \$150 million \cite{hersher2017incident}. To mitigate such defects, identifying development anti-patterns can be valuable. Anti-patterns can be defined as recurring practices associated with negative outcomes \cite{neill2011antipatterns}. These anti-patterns can be observed by analyzing open-source repositories, as they can reveal common development practices that correlate with defective IaC scripts. We refer to anti-patterns in IaC scripts that could result in excessive resource consumption cloud infrastructures as sustainability smells. In the process of developing IaC, some sustainability smells can appear from various sources, such as over-provisioning cloud resources, choosing inappropriate VM types or sizes, inefficient workload distribution, creating unnecessary network traffic, or failing to leverage energy-saving features or policies \cite{Amazon_Web_Services_2023}. Cost reduction and efficient resource utilization are among the main motivations for cloud adoption \cite{Andrikopoulos2012}. But, the user experience, quality of service, environmental footprint, and operational costs of cloud applications may all be negatively impacted by sustainability smells. However, % to the extent of our knowledge, 
empirical research on the prevalence and features of sustainability smells in IaC scripts is lacking. Our research addresses a gap in empirical studies on sustainability smells in IaC scripts, particularly in Terraform. By analyzing open-source Terraform repositories, we aim to uncover common sustainability smells that impact resource efficiency and environmental sustainability in cloud infrastructure. Specifically, we investigate the following research questions: 
\begin{itemize}
    \item \textbf{RQ1: What sustainability smells occur in IaC scripts?}
    This research question aims to identify specific patterns in IaC scripts that lead to inefficient resource utilization. By analyzing industry reports and best practices from major cloud providers such as AWS, Azure and GCP. 
    
    \item \textbf{RQ2: How do practitioners perceive sustainability smells?}\\
    This research question explores the perceptions of practitioners towards the identified sustainability smells. Through a survey from experienced cloud practitioners, we seek to understand their views on these smells and their impact. This insight is crucial for validating the practical relevance of our findings and ensuring that recommended best practices are both effective and feasible in real-world scenarios. 

    \item \textbf{RQ3: How frequently do sustainability smells occur in IaC scripts?}\\
    This question investigates the prevalence of identified sustainability smells within IaC scripts. By examining a large dataset of Terraform scripts, we aim to quantify how often these bad practices occur in real-world configurations. The frequency analysis helps highlight the commonality of these inefficiencies, providing a basis for targeted interventions and improvements in IaC practices. 
\end{itemize}

\section{Background}
In this section, we define some foundational concepts for our study, and highlight challenges that motivate our study, particularly regarding sustainability standards in IaC.

\subsection{Infrastructure as Code (IaC)}
Infrastructure as Code is an approach of managing and provisioning computing infrastructure through machine-readable definition files, rather than physical hardware configuration or interactive configuration tools \cite{10.5555/2838841}. This approach uses high-level scripting languages to automate the setup, configuration, and management of infrastructure components such as servers, networks, and databases. IaC is a key DevOps practice and an essential aspect of continuous delivery \cite{microsoft_forrester_2015}. It allows organizations to automate the provisioning process, ensuring that environments are set up consistently every time, and enabling teams to track changes, collaborate, and roll back to previous configurations if necessary \cite{gartner_2571419}. However, while IaC streamlines infrastructure management, it introduces new challenges around sustainability, particularly regarding resource optimization and environmental impact.

\subsection{Sustainability Standards in IaC}
% Terraform is an open-source IaC software tool created by HashiCorp \cite{terraform_def}. It allows users to define and provision data center infrastructure using a high-level configuration language known as HashiCorp Configuration Language (HCL), or optionally JSON \cite{terraform_configuration_syntax}. Terraform is designed to manage resources across a wide range of cloud providers as well as custom in-house solutions. The main components of Terraform include the configuration files that describe the desired state of infrastructure, the execution plans that show what actions Terraform will take to reach that state, and the state files that track the current state of resources managed by Terraform \cite{terraform_configuration_syntax}. Terraform supports multiple cloud platforms such as AWS, GCP, Azure, and many others, making it a versatile tool for managing multi-cloud infrastructure \cite{terraform_vs_chef_puppet}. Terraform enables declarative infrastructure management, where users specify what the infrastructure should look like rather than the steps to create it. This declarative approach ensures consistency and repeatability in infrastructure deployments. Terraform’s capability to manage dependencies between resources ensures that changes are applied in the correct order, reducing the risk of errors during deployment. It also supports the concept of modules, which are reusable configurations that promote best practices and reduce duplication of code \cite{turnbull_terraform_2016}.

Although there are industry standards for cloud security, compliance, and efficiency, there is no established framework specifically focused on IaC sustainability \cite{azure_security_framework, google_cloud_security_framework, aws_security_pillar}. While cloud providers like AWS, Azure and Google Cloud offer guidelines for cloud sustainability, these recommendations vary in scope and focus, leading to fragmented sustainability practices across different providers. For example, some platforms prioritize energy-efficient data center locations, while others focus on reducing idle resource usage. The lack of uniform sustainability standards for IaC results in an ad hoc approach, where practitioners independently interpret and apply sustainable practices, leading to inconsistency. Our research seeks to address this gap by exploring and proposing structured guidelines for sustainable IaC practices, encouraging more consistent and effective deployment strategies that align with sustainability objectives.

\subsection{Code Smells}
Code smells, a concept popularized by Martin Fowler in his seminal book ``Refactoring: Improving the Design of Existing Code" \cite{Fowler1999} which refer to any characteristic in the source code that suggests deeper issues, even though they do not prevent the program from functioning. These smells signal the need for refactoring to improve maintainability, readability, and quality. Code smells can arise in various forms, such as duplicated code, overly long methods, large classes, or complex inheritance hierarchies \cite{refactoring_smells}. Beyond code, similar smells can manifest in configuration files, software models, or software model transformations where inefficient structures lead to poor maintainability and potential vulnerabilities \cite{Mumtaz_2019, Panahandeh_2021}. Smells in configuration files might include hardcoded values or redundant configurations, while smells in software models can undermine maintainability, introduce redundancy, or create inefficiencies. Smells can affect various aspects of software beyond just code quality, extending into areas like security, and sustainability. When it comes to configuration files like IaC for example, security smells have been identified as potential vulnerabilities that could compromise system safety, such as hardcoded credentials, or weak access controls \cite{rahman2019seven}. These security smells serve as indicators of deeper structural problems that, if ignored, could lead to breaches or other severe security failures. In addition to security, other aspects like energy and sustainability have emerged as critical considerations, where some code smells can cause high energy consumption \cite{Bangash_2023, Gupta2024}. 

\section{Methodology}
\begin{figure*}  
    \centering
    \includegraphics[width=0.7\linewidth]{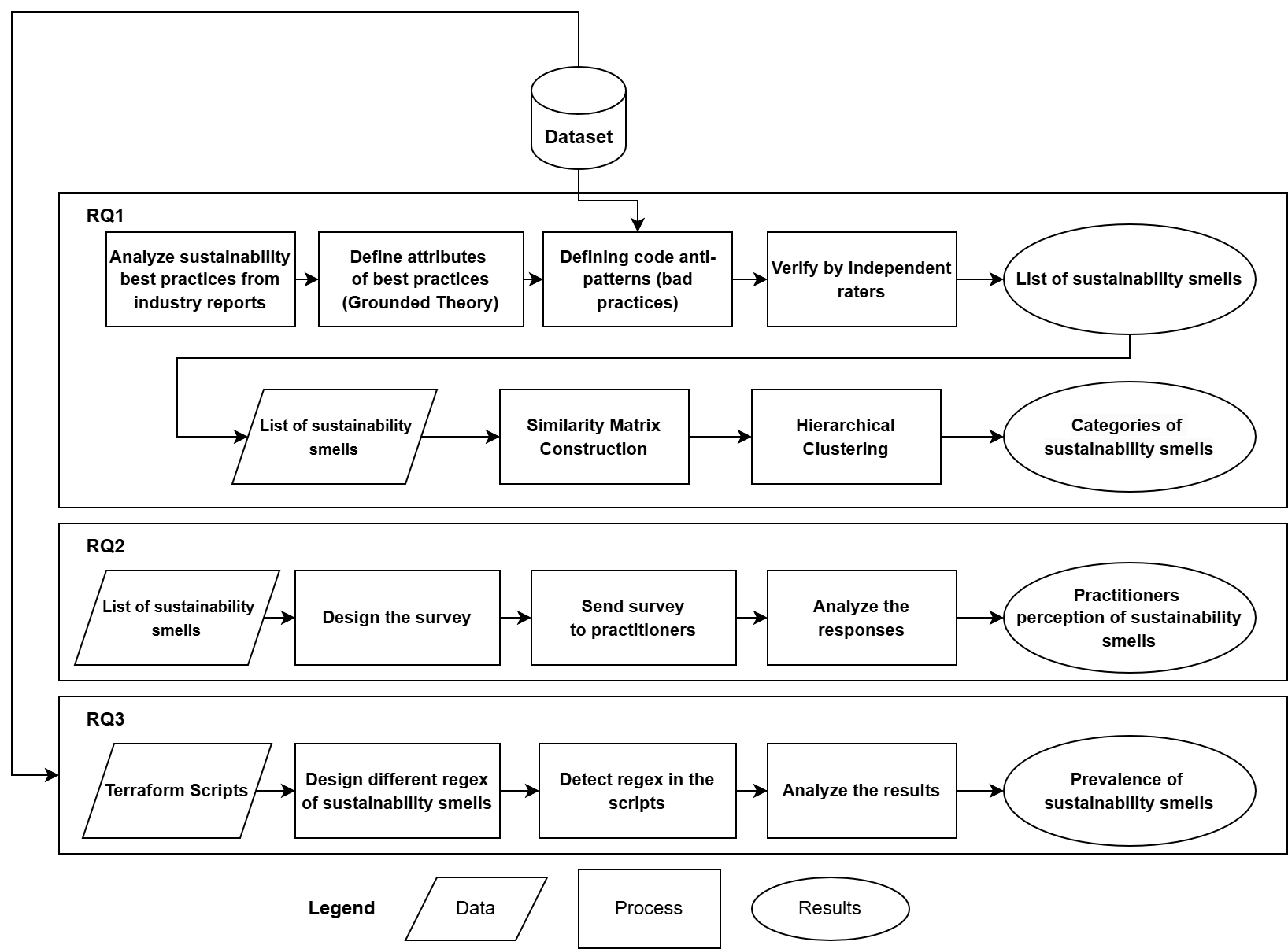}
    \caption{Overview of The Research Methodology}
    \label{methodology}
\end{figure*}

In this section, we describe the execution plan of our study. Figure \ref{methodology} shows the overview of the study methodology. In this study, we applied grounded theory to analyze a set of frameworks, identifying sustainability smells specific to IaC. We began by extracting these smells and defining their corresponding code signatures to ensure they could be empirically recognized in codebases. To validate and contextualize our findings, we conducted a survey with industry practitioners, gathering insights on the relevance and impact of these smells in real-world settings. Finally, we reported both the survey results from practitioners and empirical findings from analyzing public repositories on GitHub, providing a comprehensive view of sustainability smells in IaC.

\subsection{Analyzing Sustainability Best Practices}
To conduct a rigorous, well-structured qualitative analysis of cloud sustainability best practices, we employed Strauss and Corbin’s version of Grounded Theory \cite{corbin2014basics}. This approach was chosen due to its suitability for developing detailed categories from complex data sources and its emphasis on an iterative process between data collection and analysis. We began by selecting the sustainability reports from AWS, Azure, and GCP as primary data sources, given their comprehensive guidelines on sustainable cloud practices.

The data sources for this study were selected based on their relevance to widely used cloud platforms—AWS, Azure, and GCP—which each offer sustainability guidance cloud practices. These reports were obtained from each provider's official documentation, and they are structured around sustainable practices for cloud infrastructures, highlighting areas such as energy efficiency, resource management, and cost reduction. The AWS Well-Architected Framework's Sustainability Pillar outlines specific recommendations on regional resource selection, workload alignment, and energy-efficient architecture \cite{Amazon_Web_Services_2023}. Similarly, Azure and GCP’s sustainability best practices for Terraform emphasize optimizing cloud resources, workload scaling, and minimizing waste \cite{GCP, azure}. 

Throughout the study, we adopted an iterative process that involved cycling between data collection and analysis, following the constant comparison and theoretical sampling techniques foundational to Strauss and Corbin’s approach to Grounded Theory. Two researchers independently coded each document and then compared their codes to identify similarities and differences. Discrepancies were discussed and resolved through consensus meetings, allowing us to refine codes continuously. When new patterns or gaps emerged, we re-sampled from other sections of AWS, Azure, and GCP documents to validate findings, ensuring that emerging themes accurately captured the breadth of sustainable practices across platforms. 
% This iterative process continued until theoretical saturation was reached, where no new information modified the existing codes, confirming the robustness of the identified attributes and their relevance to sustainability smells. 

% We implement an iterative process of Open, and Axial Coding. This process enables us to explore patterns and develop a well-grounded understanding of each practice and its implications for sustainability within Terraform.

% The criteria for choosing the best practices comes from the established correlation between cloud infrastructure costs and energy consumption. Research by Horri et al. \cite{Horri2015} highlights that energy consumption in cloud environments can be modeled using cost-based models. Given this connection, our criteria for selecting the best practices were based on the premise that non-alignment with a best practice would lead to increased costs and, by extension, higher energy consumption. 

To ensure theoretical saturation, we conducted iterative rounds of open and axial coding until no new categories or codes emerged from the AWS, Azure, and GCP sustainability reports. Saturation was evidenced by the recurrence of key attributes—\textbf{Runtime Dependency}, \textbf{Resource Context}, \textbf{Code Dependency}, and \textbf{Inherent Badness}—across diverse best practices without additional variation in later cycles of analysis. This consistency indicated that our attributes sufficiently captured the variation in sustainability practices and that additional data was unlikely to yield new insights. 

% Through this rigorous Grounded Theory analysis, each attribute was naturally derived, reinforcing their validity in the context of AWS, Azure and GCP sustainability best practices. By integrating the attributes of \textbf{Runtime Dependency}, \textbf{Workload Context}, \textbf{Code Dependency}, and \textbf{Inherent Badness}, our framework offers a comprehensive and systematic approach to define the attributes of the best practices, supporting our broader research objectives.

% Finally, we analyze these groups to identify overarching themes that would form our key sustainability smells, verifying each attribute’s relevance across cloud providers. We selectively refined our final groups by choosing practices that were universally applicable. For instance, Resource Context emerged as a defining characteristic of the best practice since each provider emphasized resource right-sizing or efficient region selection. Similarly, Inherent Badness highlighted universally inefficient practices like redundant resource allocation.

\subsection{Identifying Sustainability Smells}
With a comprehensive list of best practices for sustainability derived from major cloud providers' recommendations, we can now proceed to the next stage: identifying the sustainability smells. The central premise is that each best practice extracted from AWS, Azure, or GCP has a corresponding anti-pattern that signifies a failure to adhere to these guidelines. Our objective is to leverage the dataset we have developed to uncover these anti-patterns and compile a corresponding list that reflects the identified sustainability smells. Transitioning from the established best practices and their associated attributes, we systematically explore the anti-patterns that arise from deviations from these guidelines. By examining the key sustainability attributes we defined, we can identify specific anti-patterns that correspond to the failure to implement these practices effectively. For instance, an anti-pattern associated with Runtime Dependency may involve neglecting dynamic scaling, leading to inefficient resource allocation. Once we identify these anti-patterns, we delve into our dataset to uncover specific code patterns that embody these deficiencies. This involves qualitatively analyzing a subset of the dataset we collected, meticulously reviewing the structure, resource definitions, and configuration settings of each script over eight weeks to map out how these anti-patterns manifest in actual code. By establishing this connection, we create a comprehensive framework that links best practices, their corresponding anti-patterns, and the specific code patterns that reflect these inefficiencies, ultimately guiding our understanding of sustainability smells in cloud deployments.

% The criteria for choosing the best practices comes from the established correlation between cloud infrastructure costs and energy consumption. Research by Horri et al. \cite{Horri2015} highlights that energy consumption in cloud environments can be modeled using cost-based models. Given this connection, our criteria for selecting the best practices were based on the premise that non-alignment with a best practice would lead to increased costs and, by extension, higher energy consumption. To choose the the best practices, we identified keywords that are frequently associated with cost inefficiencies, resource utilization, and energy consumption in cloud environments. These keywords include: ``overprovisioning", ``underutilization," ``inefficiency," ``idle" ``autoscaling," and ``transfers" were prioritized. By focusing on these keywords, we were able to systematically identify practices that, when violated, would contribute to increased operational costs and energy consumption, thereby validating our criteria for selecting the best practices. The selected practices from AWS include SUS01-BP01, SUS02- BP01, SUS03-BP02, and SUS04-BP05.

\subsection{Sustainability Smells Categorization}
We decided to categorize the sustainability smells we derived to better understand their different impacts and contexts, enabling more targeted and effective mitigation strategies for enhancing cloud deployment sustainability. To categorize sustainability smells credibly and systematically, we adopted a rigorous research methodology. We use hierarchical clustering because it is a powerful statistical method used to group similar data points based on their characteristics, and it works by recursively merging or splitting clusters based on their similarity, forming a tree-like structure known as a dendrogram \cite{10.5555/2823827}. We begin by reviewing the attributes of the sustainability smells we previously defined, analyzing each one to identify the specific attributes that characterize it. The results of this analysis are presented in Table II, where each row corresponds to a sustainability smell. Each cell within the table indicates whether a particular attribute is present for that smell, with a value of 1 indicating the presence of the attribute and a value of 0 indicating its absence. Then we calculate the semantic similarity between all the combinations of sustainability smells pairs. The similarity is in a form of a matrix that was populated by comparing the attributes of each smell. Based on our definition of similarity, if two smells have the same attributes, their similarity score would be 1. We used only 0 and 1, where 0 indicates no similarity and 1 indicates complete similarity. It is only 0 or 1 because the similarity between two smells is not some quantity that can be numerically quantified as it is based on a complete match of attributes between two smells. This allows for the identification of natural groupings within the data, providing insights into the relationships and patterns among different sustainability smells. This technique is particularly effective for organizing complex datasets, such as sustainability smells in IaC scripts, where there are numerous dimensions and attributes to consider.

\subsection{Gathering Practitioners' Perception of Sustainability Smells}
To understand how real-world practitioners perceive the impact of these sustainability smells, we conducted a survey targeting developers who have experience with IaC using Terraform. To validate our defined sustainability smells, we surveyed a broader population of cloud practitioners. We recruited practitioners who have experience with IaC and work at major cloud providers such as AWS, Azure, and GCP. We also included researchers working in the fields of sustainability and cloud computing to gain a broader perspective, totaling 19 survey participants. Our survey comprised of 22 questions on Google Forms, including a mix of multiple-choice and open-ended questions. The survey began with demographic questions and participant experiences with Terraform and cloud infrastructure. Subsequently, we presented participants with code snippets illustrating each sustainability smell and asked them to assess how often they encounter this pattern at their work and whether they perceive the pattern as bad practice or not (Hence a valid sustainability smell). We asked two questions per sustainability smell (a total of 14). We quantitatively analyzed the closed-ended questions to understand developers’ perceptions of the definition and examples of the shown sustainability smells.

\subsection{Dataset Collection}
We utilized a dataset meticulously gathered from GitHub repositories. The dataset collection process ensured a thorough and representative sample of real-world projects deploying cloud infrastructures. Below, we describe in detail the content and structure of this dataset, emphasizing its relevance and utility for our research. The dataset comprises a total of 812 distinct GitHub repositories containing Terraform files. These repositories are selected based on their deployment configurations for the three major cloud providers: AWS, Azure, and GCP. The dataset collection was performed over different periods for each provider to ensure an up-to-date and comprehensive set of Terraform scripts. The breakdown before data filtration is as follows:

\begin{itemize}
    \item AWS: 3,245 repositories.
    \item Azure: 1,308 repositories.
    \item GCP: 1,518 repositories.
\end{itemize}

The repositories were selected using the GitHub Code Search API, which allows for searching specific code snippets within the source files, rather than relying on metadata or file names. This method ensures that the dataset includes a wide variety of projects where Terraform is used, regardless of the primary focus of the repository. To ensure the quality and relevance of the dataset we applied several filtering criteria:

\begin{itemize}
    \item Size: The repository must have at least one file with non-zero size (i.e., $>$ 0 KB).
    \item Originality: The repository must not be a fork.
    \item Popularity: The repository must have at least 2 stars.
    \item Data Availability: The repository must be publicly accessible via the GitHub API.
    \item Content: The repository must not be a course assignment, tutorial, or intentionally insecure project.
\end{itemize}

After applying these filters, the final dataset comprises of 401 AWS repositories after removing 12 that did not meet the content criterion, 137 Azure repositories after removing 6 that did not meet the content criterion, and 274 GCP repositories after removing 7 that did not meet the content criterion.

From these repositories, we extracted all Terraform files, obtaining a total of 28,327 Terraform scripts. We sampled a subset of the dataset to facilitate a detailed qualitative analysis for identifying code patterns representing sustainability smells. For this purpose, we employed a stratified sampling approach to ensure diversity and representation across the collected repositories \cite{Botev2017}. First, we defined each repository as a stratum, as each repository may have unique characteristics that could influence sustainability practices. Within each repository, we identified Terraform scripts as the primary unit of analysis. For each stratum (repository), we randomly selected 4–5 Terraform scripts to ensure balanced representation while accounting for the variation in project complexity and configuration practices. The random selection was conducted using a random number generator to select scripts within each repository without bias. This approach allowed us to capture a diverse set of coding practices and sustainability smells while maintaining the integrity of each repository’s specific configuration approach. By selecting scripts from different repositories within each stratum, we captured a variety of coding practices and patterns, enhancing the robustness of our qualitative analysis. The result of this sampling process was a set of 1,860 Terraform scripts. This subset was thoroughly examined to uncover specific code patterns and practices associated with sustainability smells. The dataset can be found in our replication package. \footnote{Replication Package: https://doi.org/10.5281/zenodo.14020461}

\subsection{Measuring the Prevalence of Smells}
To measure the prevalence of sustainability smells, we need to know how these identified sustainability smells can be effectively detected within Terraform scripts. We developed tailored regular expression (regex) patterns based on specific attributes of each identified smell. First, we reviewed each sustainability smell and determined keywords, syntax structures, and configuration patterns that could serve as reliable indicators. Each regex pattern was iteratively refined by testing it against a subset of scripts from our dataset. After initial pattern development, we conducted a manual validation step to assess detection accuracy. A random sample of scripts detected by each pattern was manually reviewed to confirm that the detected smells aligned with the defined criteria. We further validated the regex accuracy by calculating a precision metric from these manual checks, ensuring the patterns reliably captured sustainability smells with minimal false positives. This process allowed us to validate that our regex-based method provided accurate and consistent detection across the dataset. While regex provided a feasible approach for this initial study, future work could explore more advanced detection techniques to enhance precision and scalability.

%%%%%%%%%%%%%%%%%%%%%%%%%%%%%%%%%%%%%%%%%RESULTS START%%%%%%%%%%%%%%%%%%%%%%%%%%%%%%%%%%%%%%%%%
\section{Results}
\subsection{RQ1: What sustainability smells occur in IaC scripts?}

Our analysis begins with open coding, going through all the best practices in the AWS sustainability pillar and identifying distinct phrases. These phrases were then translated into codes, for example, the phrase ``scale your infrastructure dynamically to match supply of cloud resources" coded as \textbf{Auto-Scaling} highlighting a practice that encourages auto-scaling resources. Similarly, ``Understand the devices and equipment used in your architecture and use strategies to reduce their usage” was coded as \textbf{Minimize Hardware Amount} marking potential workload context dependencies for sustainability. To ensure comprehensive coverage, we also reviewed sustainability recommendations from other major cloud providers following the same criteria, such as GCP \cite{GCP} and Azure \cite{azure}. See Table \ref{codes_examples} for examples of all the codes derived.

\begin{table*}[hbt!]
\centering
\caption{Example Codes Derived from AWS, Azure, and GCP Sustainability Reports}
\label{codes_examples}
\begin{tabular}{|l|l|p{5cm}|p{5cm}|}
\hline
\textbf{Code} & \textbf{Source Report} & \textbf{Example Sentence 1} & \textbf{Example Sentence 2} \\ \hline
Time Consuming Resources & AWS, Azure & 
\textbf{AWS:} ``Optimize your code that runs within different components of your architecture to minimize resource usage while maximizing performance". & 
\textbf{Azure:} ``Optimize component costs. Regularly remove or optimize legacy, unneeded, and underutilized workload components, including application features, platform features, and resources". \\ \hline

Auto-Scaling & AWS, Azure & 
\textbf{AWS:} ``Use elasticity of the cloud and scale your infrastructure dynamically to match supply of cloud resources to demand and avoid overprovisioned capacity in your workload". &
\textbf{Azure:} ``Optimize scaling costs. Evaluate alternative scaling within your scale units". \\ \hline

Remove Low Usage Resources & AWS, GCP & 
\textbf{AWS:} ``Use efficient software and architecture patterns such as queue-driven to maintain consistent high utilization of deployed resources". &
\textbf{GCP:} ``Idle resources incur unnecessary costs and emissions". \\ \hline

Minimize Hardware Amount & AWS, GCP & 
\textbf{AWS:} ``Use the minimum amount of hardware for your workload to efficiently meet your business needs". &
\textbf{GCP:} ``Consider serverless options for workloads that don't need VMs. These managed services often optimize resource usage automatically, reducing costs and carbon footprint". \\ \hline

Optimize Hardware Accelerators & AWS & 
\textbf{AWS:} ``Optimize your use of accelerated computing instances to reduce the physical infrastructure demands of your workload". & 
\textbf{N/A}
\\ \hline

Minimize Network Traffic & AWS, GCP & 
\textbf{AWS:} ``Use shared file systems or object storage to access common data and minimize the total networking resources required to support data movement for your workload". &
\textbf{GCP:} ``minimize the total networking resources required to support data movement for your workload" \\ \hline

Unnecessary Backups & AWS & 
\textbf{AWS:} ``Avoid backing up data that has no business value to minimize storage resources requirements for your workload".&
\textbf{N/A}\\ \hline

Upgrade for sustainability & AWS, Azure & 
\textbf{AWS:} ``Continually monitor and use new instance types to take advantage of energy efficiency improvements".& 
\textbf{Azure:} ``Optimize environment costs. Align spending to prioritize preproduction, production, operations, and disaster recovery environments. For each environment, consider the required availability, licensing, operating hours and conditions, and security". \\ \hline

Redundant data & AWS & 
\textbf{AWS:} ``Remove unneeded or redundant data to minimize the storage resources required to store your datasets".&
\textbf{N/A}\\ \hline

State Management & GCP & 
\textbf{GCP:} ``The Terraform state file is critical for maintaining the mapping between Terraform configuration and Google Cloud resources. Corruption can lead to major infrastructure problems".&
\textbf{N/A}\\ \hline

\end{tabular}
\end{table*}

During open coding, we observed that sustainability best practices for cloud infrastructures can vary significantly. A best practice suitable for one type of infrastructure might be irrelevant or inefficient for another. For instance, in a known infrastructure setup, named Infrastructure \textbf{A}, where storage demands are stable and constant (e.g., always needing X storage), practices such as auto-scaling based on demand may be unnecessary. Conversely, for another Infrastructure \textbf{B}, where storage requirements fluctuate depending on workload, implementing demand-based resource scaling becomes essential to manage variability and avoid inefficiencies. This distinction highlights the importance of tailoring sustainability strategies to the specific needs and operational behaviors of each infrastructure type. Observing variability in sustainability best practices across cloud infrastructures highlighted the need to define specific attributes for these practices. These attributes help explain the contextual factors of an infrastructure that make a given best practice applicable. For example, Infrastructure \textbf{A} might have constant storage needs, whereas Infrastructure \textbf{B} experiences demand fluctuations, making certain practices like auto-scaling relevant only to \textbf{B}. This idea led to axial coding, where we defined attributes to capture these factors systematically, thereby identifying when and where sustainability practices are most effectively applied.

In the axial coding phase, we focused on identifying commonalities among initial codes generated during open coding to derive higher-level themes, or ``attributes," that characterize sustainability best practices across cloud platforms. To start, we examined each code's relationship to broader sustainability goals, analyzing if it was influenced by runtime workloads, resource configurations, application code, or if it represented a universally inefficient practice. For example, practices like ``SUS02-BP01 Scale workload infrastructure dynamically” from AWS were linked with \textbf{Runtime Dependency} as they are designed to dynamically adjust resource allocation in response to workload fluctuations. Similarly, the codes of the best practices ``CO:07	Optimize component costs” and ``Choose the most suitable cloud regions" from Azure and GCP were assigned to a category indicating specific infrastructure configurations, leading to the \textbf{Workload Context} attribute. For another attribute we formed which is \textbf{Code Dependency}, we examined codes that are closely tied to the application's structure or software requirements. For instance, AWS's ``SUS04-BP07 Minimize data movement across networks" hints at the need to minimize the total networking resources required to support data movement for the workload (which is controlled by the actual code of the running application on the cloud) pointing toward a dependency on the code structure rather than on infrastructure alone. Finally, we grouped universally inefficient practices, such as idle resources or excessive logging, under \textbf{Inherent Badness}, as these practices are considered bad for sustainability regardless of workload context.

After examining the dataset, we identified a set of sustainability smells that violate the best practices. We will refer to sustainability smells from now on with the notation (SSx) where x is the number of the smell. The identified sustainability smells include the following:

\begin{itemize}
    \item \textbf{SS1: Over-Provisioning Resources} \\ 
    Over-provisioning in Terraform scripts involves allocating more resources than necessary, often leading to wasted capacity. This can be avoided by regularly reviewing resource usage and scaling down where possible to match actual needs \cite{Amazon_Web_Services_2023}.

    \item \textbf{SS2: Lack of Auto-Scaling} \\
    A lack of auto-scaling in Terraform scripts results in static resource allocation, disregarding workload fluctuations. Implementing auto-scaling mechanisms helps dynamically adjust resources to meet demand efficiently \cite{azure}.

    \item \textbf{SS3: Ignoring Resource Lifecycles} \\
    Failing to manage resource lifecycles leads to inefficient resource usage and higher costs. Establishing lifecycle policies, such as create\_before\_destroy or prevent\_destroy, ensures resources are efficiently managed over time \cite{Amazon_Web_Services_2023}.

    \item \textbf{SS4: Excessive Logging} \\
    Enabling detailed logging for all resources without considering necessity increases storage and processing costs. To optimize, set retention policies that limit logs to essential data and required retention periods \cite{Amazon_Web_Services_2023}.

    \item \textbf{SS5: Unoptimized Data Transfers} \\
    Unoptimized data transfers, where resources communicate across regions, lead to unnecessary costs and energy use. To mitigate, co-locate frequently interacting resources in the same region \cite{Amazon_Web_Services_2023}.

    \item \textbf{SS6: State Management} \\
    Robust state management via remote backends supports centralized infrastructure control and minimizes errors. Using remote backends enhances consistency and supports sustainable practices \cite{GCP}.

    \item \textbf{SS7: Monolithic Infrastructure} \\
    Monolithic configurations, where all resources are managed in a single file, reduce modularity and hinder scalability. Dividing configurations into modular components enhances flexibility and efficiency \cite{GCP}.
    
\end{itemize}

To validate our identified sustainability smells, we engaged two independent raters who were not involved in this study. These raters, possessing expertise in IaC, were tasked with assessing whether the identified smells corresponded to the associated best practices outlined in industry reports. For this purpose, we supplied the raters with the names of the smells, an example of each, and the corresponding best practice. Each rater independently evaluated the alignment of the smells with the provided best practices. Both raters consistently associated each of the seven sustainability smells with the same best practice. The evaluation yielded a Cohen’s Kappa score of 1.0, reflecting perfect agreement between the raters. \\

\begin{table*}
  \caption{Defined Sustainability Smells}
  \label{commands}
  \centering
  
  \begin{tabular}{ | l | l | l | p{5cm} |}
    \hline
    Sustainability Smell (SS) & Category  & Code Sample \\ \hline
    SS1: Over-Provisioning Resources & 2 & 
    \begin{lstlisting}
    resource "azurerm_virtual_machine" "inefficient_vm" {
        name                = "example-vm"
        location            = "East US"
        vm_size             = "Standard_D16s_v3" # Overprovisioned 
    }
    \end{lstlisting} \\ \hline

    SS2: Lack of Auto-Scaling & 2 & 
    \begin{lstlisting}
    resource "aws_instance" "app" {
        count = 5 # Fixed number of instances
        ami = "ami-0c55b159cbfafe1f0"
        instance\_type = "m5.2xlarge"
    }
    \end{lstlisting} \\ \hline
    
    SS3: Ignoring Resource Lifecycles & 1 & 
    \begin{lstlisting}
    resource "azurerm_managed_disk" "example" {
        name                 = "example-disk"
        storage_account_type = "Standard_LRS"
        create_option        = "Empty"
        
        lifecycle {
            create_before_destroy = true
        }
    }
    \end{lstlisting} \\ \hline
    
    SS4: Excessive Logging & 1 & 
    \begin{lstlisting}
    resource "aws_cloudwatch_log_group" "detailed_logs" {
        name = "detailed-log-group"
        retention_in_days = 365 # Long period for detailed logs
        ...
    }
    \end{lstlisting} \\ \hline
    
    SS5: Unoptimized Data Transfers & 3 & 
    \begin{lstlisting}
    resource "google_compute_instance" "example" {
        name         = "example-instance"
        machine_type = "n1-standard-1"
        zone         = "us-west1-a" 
        #Data frequently transferred to another region
    }
    \end{lstlisting} \\ \hline
    
    SS6: State Management & 1 & 
    \begin{lstlisting}
    terraform {
        backend "gcs" {
            bucket = "my-terraform-state"
            prefix = "network"
        }
    }
    \end{lstlisting} \\ \hline

    SS7: Monolithic Infrastructure & 1 & 
    \begin{lstlisting}
    resource "google_compute_network" "main" { ... }
    resource "google_compute_subnetwork" "example" { ... }
    # Many more resources follow...
    \end{lstlisting} \\ \hline

    \end{tabular}
\end{table*}

\begin{table*}[h]
  \caption{Attributes of Sustainability Smells}
  \label{sus_attributes}
  \centering
  \begin{tabular}{|p{5cm}|c|c|c|c|}
    \hline
    Sustainability smell (SS) & Runtime Dependency  & Resource Context & Code Dependency & Inherent Badness \\ \hline
    SS1: Over-Provisioning Resources & 1 & 1 &  0 & 0 \\ \hline
    SS2: Lack of Auto-Scaling & 1 & 1 &  0 & 0 \\ \hline
    SS3: Ignoring Resource Lifecycles & 0 & 0 & 0 & 1 \\ \hline
    SS4: Excessive Logging & 0 & 0 & 0 & 1 \\ \hline
    SS5: Unoptimized Data Transfers & 0 & 0 & 1 & 0 \\ \hline
    SS6: State Management & 0 & 0 & 0 & 1 \\ \hline
    SS7: Monolithic Infrastructure & 0 & 0 & 0 & 1 \\ \hline
    
    \end{tabular}
\end{table*}

\begin{figure}
    \centering
    \includegraphics[width=0.6\linewidth]{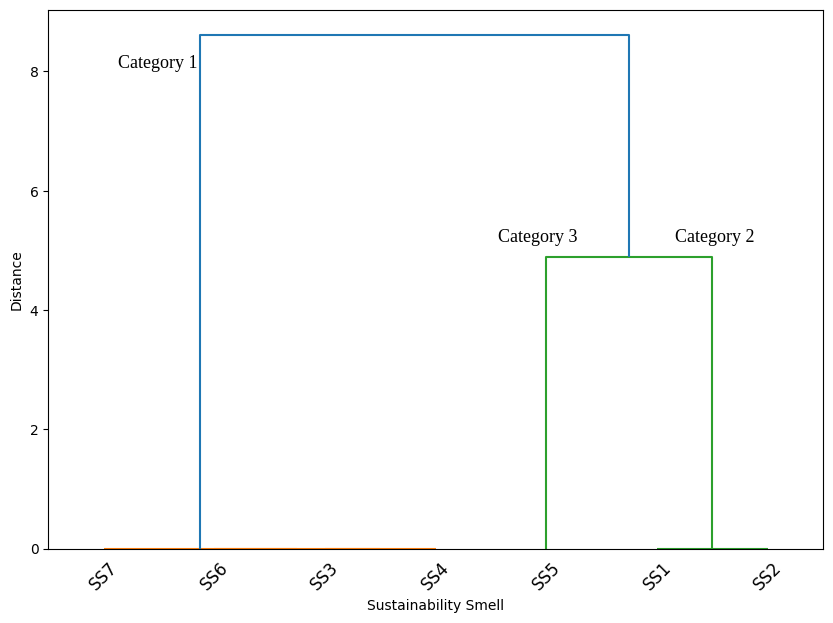}
    \caption{Hierarchical Clustering Dendrogram for Sustainability Smells}
    \label{Dendrogram}
\end{figure}

% These attributes included:
% \begin{itemize}
%     \item Runtime Dependency: Indicates whether the best practice depends on the actual runtime workload.
%     \item Resource Context: Whether the best practice depends on the context of specific resources (e.g., VM types, storage options).
%     \item Code Dependency: Whether the best practice requires knowledge of the application code being run.
%     \item Inherent Badness: Indicates whether the sustainability smell is inherently bad regardless of context.
% \end{itemize}

For each sustainability smell, we use the defined set of sustainability best practices attributes to help describe its characteristics. To ensure clarity and transparency, we organized the attributes and their corresponding values (1 if the attribute exist in the smell and 0 otherwise) for each sustainability smell in a table. An example is shown in Table \ref{sus_attributes}. We create a similarity matrix to quantify the similarity between each pair of sustainability smells. The matrix is shown below, each row and column in the matrix represents a specific sustainability smell. The entries (0 or 1) indicate whether a pair of smells shares significant similarities (1) or not (0). The similarity matrix between each pair of sustainability smells is shown below:

\begin{equation}
    \begin{bmatrix}
    1 & 1 & 0 & 0 & 0 & 0 & 0 \\
    1 & 1 & 0 & 0 & 0 & 0 & 0 \\
    0 & 0 & 1 & 1 & 0 & 1 & 1 \\
    0 & 0 & 1 & 1 & 0 & 1 & 1 \\
    0 & 0 & 0 & 0 & 1 & 0 & 0 \\
    0 & 0 & 1 & 1 & 0 & 1 & 1 \\
    0 & 0 & 1 & 1 & 0 & 1 & 1 \\
    \end{bmatrix}
\end{equation}

We use agglomerative hierarchical clustering, a bottom-up approach where each data point starts in its own cluster, and pairs of clusters are merged as we move up the hierarchy \cite{Nielsen2016}. This method is particularly suited for our needs as it does not require us to predefine the number of clusters. We constructed a dendrogram to visualize the clustering process. The dendrogram illustrates the nested grouping of sustainability smells based on their similarities. At the bottom of the dendrogram, each leaf represents an individual sustainability smell. As we move up, nodes represent clusters formed by merging pairs of smells or clusters. Based on the dendrogram we can interpret the clustering results as follows:

\begin{itemize}
    \item \textbf{Category 1: General Sustainability Smells:} Can be considered the rule of thumb (SS3, SS4, SS6, SS7).
    \item \textbf{Category 2: Demand Sustainability Smells:} Requires data about the actual demand of resources for the workload whether this comes on runtime or based on predefined requirements so that the recommendations can make sense (SS1, SS2).
    \item \textbf{Category 3: Application Sustainability Smells:} Requires data about the actual code the infrastructure will run (SS5).
\end{itemize}

\subsection{RQ2: How do practitioners perceive sustainability smells?}
\begin{figure}
    \centering
    \includegraphics[width=0.65\linewidth]{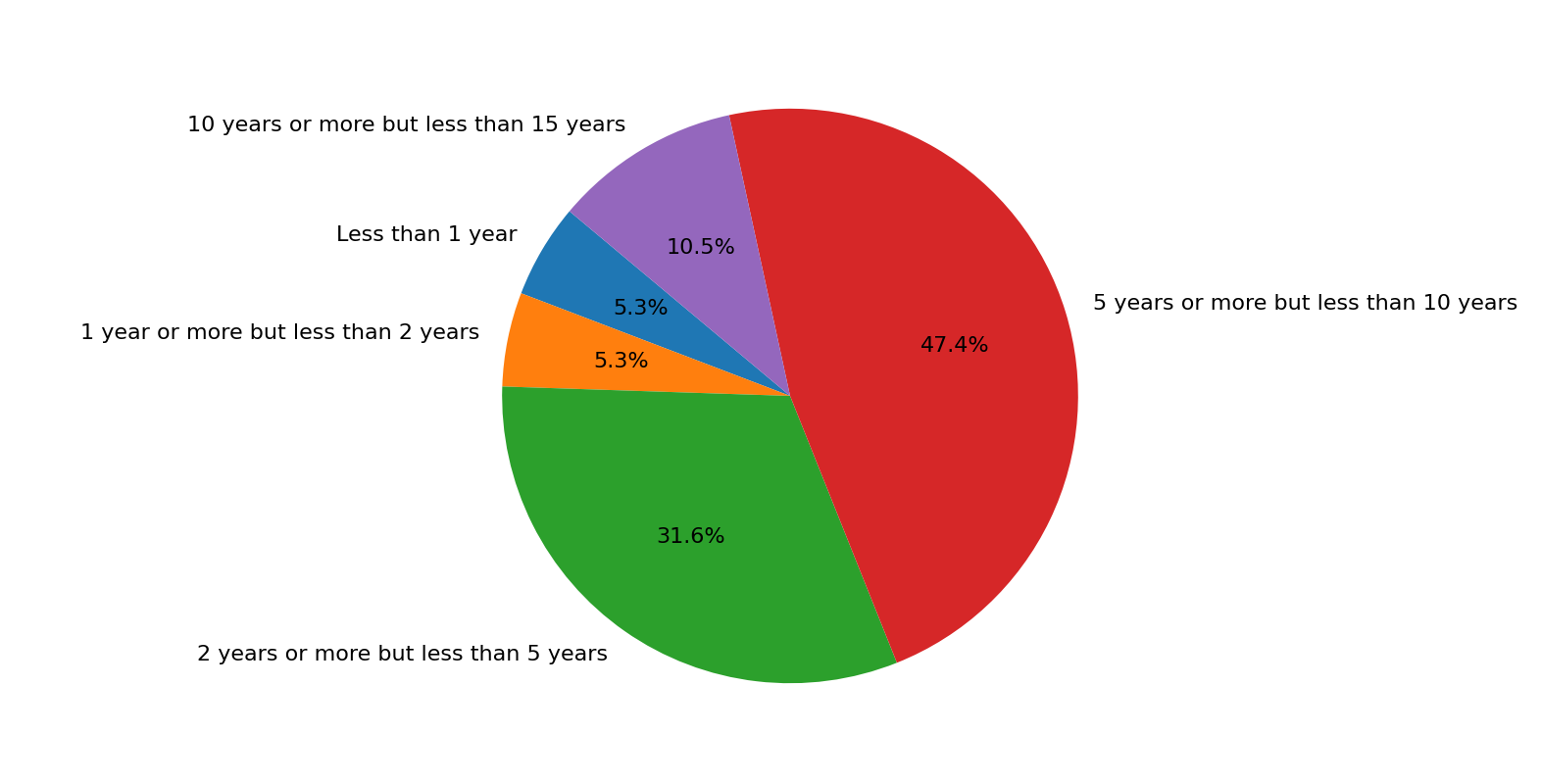}
    \caption{Survey participants YOE using IaC tools}
    \label{iac_exp}
\end{figure}

Most of the survey participants have substantial experience in programming, with 5 to 10 years in general development and an equal range of experience specifically in Infrastructure as Code (IaC). Figure \ref{iac_exp} show the IaC years of experience (YOE) for the survey participants.

We found that \textbf{SS1} and \textbf{SS7} are the most frequently encountered smells, with 52.6\% of the participants often or always facing these issues. \textbf{SS3} and \textbf{SS2} are also common, with 36.8\% of participants reporting that they encounter them sometimes. Around 26.3\% of the participants often encounter \textbf{SS2}, while 21.1\% often encounter \textbf{SS3}. \textbf{SS4} and \textbf{SS5} are less frequently encountered but still significant, with more than a third of the participants encountering these sometimes or often. \textbf{SS6} practices are another notable issue, with 31.6\% of participants facing it often and 26.3\% always facing it. Our replication package contains all the detailed responses for all the identified smells.

\begin{figure}
    \centering
    \includegraphics[width=0.75\linewidth]{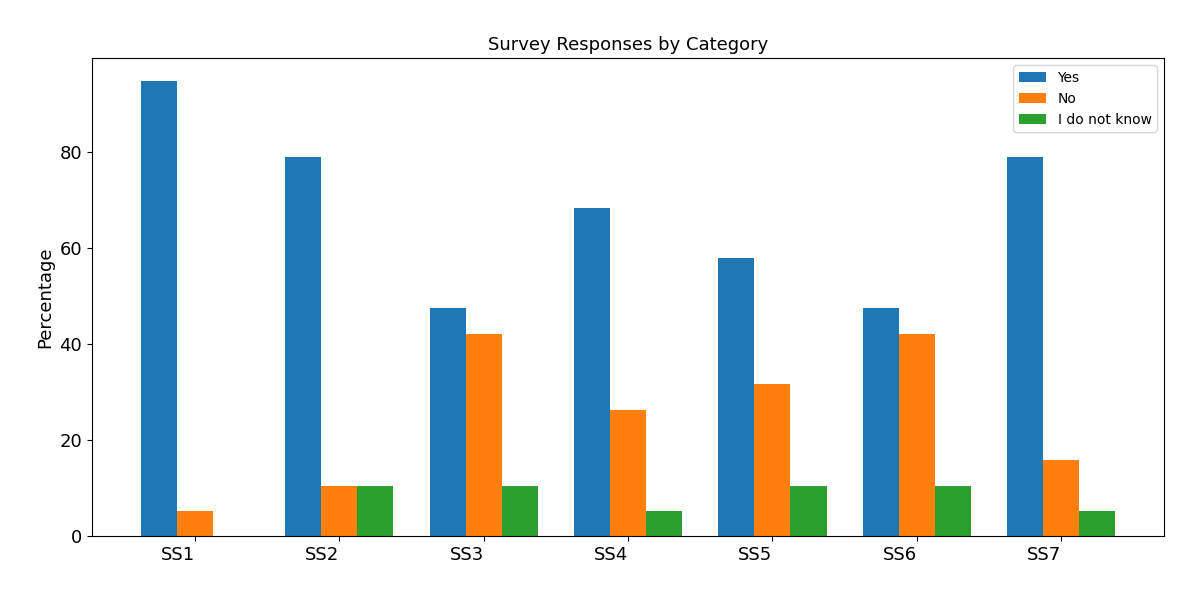}
    \caption{Survey Participants Perception of Sustainability Smells}
    \label{smell_perception}
\end{figure}

Figure \ref{smell_perception} shows the participants' perception of sustainability smells being actually bad for sustainability, it is an answer to the question ``Do you perceive this pattern as a bad practice?". The data indicates that several identified sustainability smells in Terraform received a majority vote, signaling consensus that they are generally considered bad practices. Specifically, the following percentages of participants agreed that each sustainability smell represents a negative practice: \textbf{SS1} (94.7\%), \textbf{SS2} (78.9\%), \textbf{SS3} (47.4\%), \textbf{SS4} (68.4\%), \textbf{SS5} (57.9\%), \textbf{SS6} (47.4\%), and \textbf{SS7} (78.9\%). These results highlight the prevalence of certain practices viewed as unfavorable, particularly for \textbf{SS1}, \textbf{SS2}, and \textbf{SS7}, which received notably high levels of agreement. The implications of the results can be found in our discussion in Section V.

\subsection{RQ3: How frequently do sustainability smells occur in IaC scripts?}
\begin{figure}[h]
    \centering
    \includegraphics[width=0.5\linewidth]{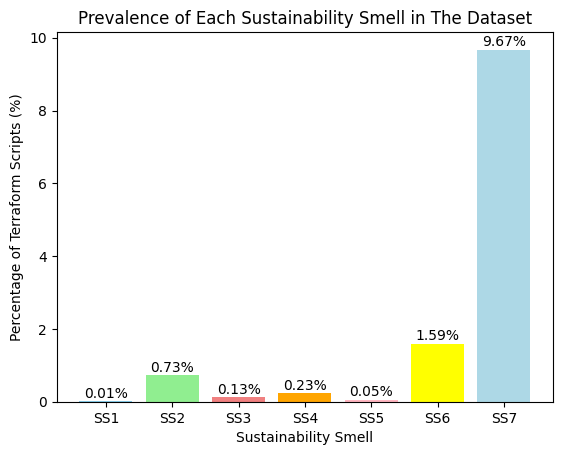}
    \caption{Percentage of Terraform Files with Specific Sustainability Smells}
    \label{rq3}
\end{figure}

Figure \ref{rq3} visually represents the prevalence of the sustainability smells across the dataset. It highlights the frequency with which each smell appears. The analysis of sustainability smells in the dataset shows that SS7 (Monolithic Infrastructure) is the most prevalent, found in 9.67\% of Terraform scripts, indicating a widespread issue with non-modular architectures. SS6 (State Management) is the second most common at 1.59\%, underscoring the challenges in managing state files effectively. Other smells, like SS2 (Lack of Auto-Scaling) at 0.73\% and SS3 (Ignoring Resource Lifecycles) at 0.13\%, occur less frequently but still point to areas needing improvement. SS4 (Excessive Logging) and SS5 (Unoptimized Data Transfers) are rare, with prevalence rates of 0.23\% and 0.05\%, respectively, while SS1 (Over-Provisioning Resources) is almost non-existent at 0.01\%. These results highlight the need for better modularization, state management, and adherence to best practices to enhance cloud deployment sustainability. Our regex-based method achieved 100\% accuracy based on the random sampling manual validation.

%%%%%%%%%%%%%%%%%%%%%%%%%%%%%%%%%%%%%%%%%RESULTS END%%%%%%%%%%%%%%%%%%%%%%%%%%%%%%%%%%%%%%%%%

%%%%%%%%%%%%%%%%%%%%%%%%%%%%%%%%%%%%%%%%%DISCUSSION START%%%%%%%%%%%%%%%%%%%%%%%%%%%%%%%%%%%%%%%%%
\section{Discussion}

The analysis of sustainability smells in IaC reveals practical recommendations for practitioners to improve efficiency and sustainability in Terraform-managed environments. To address \textbf{SS1} (Over-Provisioning Resources), practitioners should regularly review resource usage and adjust scaling to match actual needs, reducing unnecessary waste. For \textbf{SS2} (Lack of Auto-Scaling), configuring auto-scaling policies can help ensure resources dynamically adjust to workload changes. Implementing lifecycle policies (\textbf{SS3}) is recommended to manage resources effectively and avoid redundant creation and destruction. For \textbf{SS4} (Excessive Logging), setting targeted retention policies minimizes storage costs by retaining only essential logs. Co-locating resources that frequently interact (\textbf{SS5}) within the same region can reduce energy use from data transfers. Practitioners can improve \textbf{SS6} (State Management) by utilizing remote backends for consistent state tracking and management. Finally, for \textbf{SS7} (Monolithic Infrastructure), adopting a modular approach by breaking down configurations into smaller, manageable components can enhance scalability and resource efficiency. By adopting these strategies, practitioners can effectively address sustainability smells in IaC, fostering more cost-effective and environmentally sustainable cloud deployments.

The dendrogram in Figure \ref{Dendrogram} groups these smells by attributes, showing clusters of general issues (e.g., \textbf{SS3}, \textbf{SS4}, \textbf{SS6}), demand-related issues (e.g., \textbf{SS1}, \textbf{SS2}), and application-specific issues (e.g., \textbf{SS5}). These clusters indicate that effective solutions require context-specific strategies: modular design for general smells, dynamic provisioning for demand-based issues, and data management for application-specific needs. This classification provides a structured approach for improving IaC sustainability.

These survey results suggest that sustainability in IaC requires greater emphasis on automated resource management, modular design, and efficient data handling. By addressing these common smells, practitioners can create IaC scripts that are not only more efficient but also align better with sustainability goals. The detailed responses in our replication package further emphasize the need for best practices in IaC to be continuously refined and adopted by the community to achieve sustainable cloud infrastructure management.

%%%%%%%%%%%%%%%%%%%%%%%%%%%%%%%%%%%%%%%%%DISCUSSION END%%%%%%%%%%%%%%%%%%%%%%%%%%%%%%%%%%%%%%%%%

%%%%%%%%%%%%%%%%%%%%%%%%%%%%%%%%%%%%%%%%%RW START%%%%%%%%%%%%%%%%%%%%%%%%%%%%%%%%%%%%%%%%%
\section{Related Work}
Sustainable software engineering is a growing field of research, covering a broad spectrum of themes, including software code smells, software sustainability, and IaC. McGuire et al. \cite{mcguire2023sustainability} proposed a theory highlighting the layered nature of sustainable practices in software engineering. They argue that sustainability is a stratified concept with multiple layers of implementation and impact, ranging from individual software components to entire systems. Saputri and Lee \cite{saputri2021integrated} presented an integrated framework that incorporates sustainability design into the software engineering life cycle. Their empirical study shows how systematically embedding sustainability considerations can improve the overall efficiency and sustainability of software projects.

Several studies focused on the energy consumption of software and machine learning models. Castaño et al. \cite{castano2023exploring} investigated the carbon footprint of Hugging Face’s machine learning models. Hort et al. \cite{hort2023exploratory} examined the energy usage of language models for source code. Castanyer et al. \cite{castanyer2024design} explored design decisions in AI-enabled mobile applications that contribute to greener AI. Rani et al. \cite{rani2024energy} identified energy patterns in web development that lead to more efficient applications. However, we focused on the energy consumption of IaC.

In IaC, several studies analyze smells, particularly related to security. Saavedra and Ferreira \cite{saavedra2022glitch} introduced Glitch, an automated tool for detecting security smells in polyglot IaC scripts, highlighting common vulnerabilities and providing remediation strategies. Rahman et al. \cite{rahman2021security, rahman2019seven} identified and replicated studies on security smells in Ansible and Chef scripts, analyzing their prevalence and impact. These works primarily focus on identifying and mitigating security risks in IaC scripts. Meanwhile, our work look at IaC from the sustainability prospective.

Other studies focused on software quality metrics and cost awareness in IaC. Dalla Palma et al. \cite{dalla2020toward} worked towards a catalog of software quality metrics for infrastructure code. Siddiq et al. \cite{siddiq2022empirical} conducted an empirical study of code smells in transformer-based code generation techniques. Feitosa et al. \cite{feitosa2024mining} explored cost awareness in IaC artifacts of cloud-based applications. 

Our work takes a different view by focusing on the identification and categorization of sustainability smells within IaC, emphasizing the significance of computational %and energy% 
efficiency. By systematically analyzing and defining these sustainability smells based on established best practices and their cost implications, our research bridges the gap between software engineering practices and sustainable cloud infrastructure. In addition to the prior studies that concentrate on energy consumption, security, or software quality metrics, our approach comprehensively addresses the interrelated aspects of cost, energy efficiency, and adherence to best practices.
%%%%%%%%%%%%%%%%%%%%%%%%%%%%%%%%%%%%%%%%%RW END%%%%%%%%%%%%%%%%%%%%%%%%%%%%%%%%%%%%%%%%%

%%%%%%%%%%%%%%%%%%%%%%%%%%%%%%%%%%%%%%%%%TFV START%%%%%%%%%%%%%%%%%%%%%%%%%%%%%%%%%%%%%%%%%
\section{Threats to Validity}
\subsection{Internal Validity}
There is a risk of selection bias in the gathering of Terraform scripts from various online platforms. Scripts chosen for analysis may not be representative of the entire population of Terraform scripts, potentially leading to skewed results. To mitigate this, a diverse and random sampling strategy will be employed. Also, the industry reports collected may introduce bias if the selected sources predominantly focus on specific cloud platforms or industries. This bias could impact the identification of energy-intensive practices. To minimize this, a broad range of industry reports from different major cloud providers will be considered, and efforts will be made to encompass various perspectives.

\subsection{External Validity}
The study’s findings may not be universally applicable, as they are based on a specific set of Terraform scripts and cloud platforms. This poses a threat to external validity, and the study will transparently communicate its context and limitations, emphasizing its applicability to Terraform scripts and specified cloud platforms. Additionally, cloud environments are dynamic, and the identified patterns may become obsolete or evolve. The study will acknowledge this dynamic nature and emphasize the temporal relevance of its findings. The effectiveness of detection techniques may vary based on specific patterns and chosen scripts, and conclusions about performance might not be generalizable to all sustainability smells or IaC scripts. This limitation will be acknowledged, providing insights into the strengths and weaknesses of the techniques within the defined scope. Moreover, the identified code patterns may not encompass all potential manifestations of sustainability smells, as variations in coding styles, project requirements, and deployment scenarios could introduce alternative patterns. Additionally, the sustainability smells defined in this study serve as recommendations for a sustainable, cost-effective cloud infrastructures. However, removing these smells may not necessarily result in significant reductions in energy consumption or cost. The approach remains iterative and adaptive, aiming to refine and expand understanding of sustainability-related code patterns in Terraform scripts.
%%%%%%%%%%%%%%%%%%%%%%%%%%%%%%%%%%%%%%%%%TFV END%%%%%%%%%%%%%%%%%%%%%%%%%%%%%%%%%%%%%%%%%

%%%%%%%%%%%%%%%%%%%%%%%%%%%%%%%%%%%%%%%%%CONCLUSION START%%%%%%%%%%%%%%%%%%%%%%%%%%%%%%%%%%%%%%%%%
\section{Conclusion}
This study systematically identified and analyzed sustainability smells in Terraform scripts to improve cloud infrastructure efficiency. By collecting 28,327 scripts and using stratified sampling, we conducted a detailed qualitative analysis on 1,860 scripts, identifying code patterns that violate sustainability best practices. We identified sustainability smells on the AWS, GCP, and Azure sustainability recommendation. 

We conducted a survey and the responses validated these smells, noting their negative impact on cost and energy efficiency. Opinions on SS3 (Ignoring Resource Lifecycles) and SS6 (State Management) showed the complexity of these issues and the need for context-specific solutions. Quantitative analysis revealed that SS7 (Monolithic Infrastructure) was the most common smell, present in 9.67\% of scripts, followed by SS6 (State Management) at 1.59\% and SS2 (Lack of Auto-Scaling) at 0.73\%. These findings highlight key areas for sustainability improvement in cloud deployments. Our findings offer valuable insights for cloud practitioners and policymakers aiming to optimize cloud infrastructure for both cost-effectiveness and environmental sustainability. 

\bibliographystyle{unsrt}  


\end{document}